\documentstyle[preprint,aps,version2]{revtex}
\begin{document}

\begin{title}
{Asymptotically exact solution of the multi-channel resonant-level model}
\end{title}

\author{Guang-Ming Zhang\cite{gmz}, Zhao-Bin Su\cite{sy}, and Lu Yu\cite{sy}}
\begin{instit}
 {International Center for Theoretical Physics, P. O. Box 586,
34100 Trieste, Italy}
\end{instit}

\begin{abstract}
An asymptotically exact partition function of the multi-channel resonant-level
model is obtained through Tomonaga-Luttinger bosonization. A Fermi-liquid vs
non-Fermi-liquid transition, resulting from a competition between the Kondo
and X-ray edge physics, is elucidated explicitly via the renormalization
group theory. In the strong-coupling limit, the model is renormalized
to the Toulouse limit.
\end{abstract}

\newpage
%\narrowtext
The central issue under current debate is whether one should involve
non-Fermi liquid (FL) behavior to interpret the normal state properties of
the high-$T_c$ superconductors and whether such a behavior can be derived
from models of strong electron correlations [1,2]. Except for one
dimension, perturbation treatments can not provide any mechanism for the
breakdown of FL behavior [3]. Thus, it is of great interest to study models
whose properties can be obtained either exactly or within controllable
approximations, exhibiting a transition from FL to non-FL behavior as some
parameters of the models are varied.

Unfortunately, lattice models have so far proven intractable to show such a
transition. On the other hand, some single-impurity models, for example,
single-channel Anderson or Kondo model, show a local FL behavior [4], whereas
the others, {\it e.g.}, the X-ray edge [5] or the multi-channel
Kondo problem [6,7] exhibit non-FL characteristics. Recently, Varma {\it et
al.} have explored the possible FL--non-FL transitions in  the generalized
Anderson model [8] and its spinless version: multi-channel resonant-level
(MCRL) model [9,10], which should shed some light on similar transitions
in lattice models. Another type of generalized Anderson model has been also
considered in the context of infinite dimensions [11].

The single-channel resonant-level model can be derived from the Kondo
model through Tomonaga-Luttinger bosonization and it exhibits the same
local FL behavior below the Kondo temperature as for the Anderson
impurity model [12,13]. The generalization of this model to include finite
range Coulomb repulsion leads to a coupling to more orbital channels in
addition to the hybridizing channel itself [9]. The spinless version of
this generalized Anderson impurity model may contain the basic physics of
the FL vs non-FL transition in the symmetric case [8].

In this Rapid Communication, we first use bosonization [14] to derive the
asymptotically
exact partition function for the MCRL model. Then, we apply the poor-man's
scaling theory [15] to calculate the renormalization group (RG) equations
from which the physics of a FL--non-FL transition is
explicitly demonstrated. In fact, we can separate the model Hamiltonian into
"hybridizing" and "screening" parts. The first part is just
the usual single-channel resonant-level model describing the Kondo physics,
exhibiting a local FL behavior [12,13], while the second
part is the multi-channel X-ray edge model describing the Anderson
orthogonality
catastrophe [5], displaying a local non-FL behavior. The
essential physics of the MCRL model is thus the
competition between the FL-type Kondo resonance and the non-FL-type
Anderson catastrophe. When the screening part has at least three channels
and strong enough repulsive interactions with impurity, the
model displays the X-ray edge singularities; otherwise, the model is
describing the Kondo physics of the hybridizing part. There exists a
transition between these FL and non-FL metallic phases, mathematically
similar but physically different from the usual
Kosterlitz-Thouless phase transition corresponding to a
metal-insulator transition [16]. The transition concerned here is
just a crossover behavior rather than a genuine phase transition.

The MCRL model Hamiltonian is given by [10]
%\widetext
\begin{equation}
H= \sum_{k,l}\epsilon_k C_{k,l}^{\dag}C_{k,l} + \frac{t}{\sqrt{N}}
  \sum_k (C^{\dag}_{k,0}d + d^{\dag}C_{k,0}) + \frac{1}{N}\sum_{k,k',l}
  V_l C^{\dag}_{k,l}C_{k',l}( d^{\dag}d-\frac{1}{2}),
\end{equation}
%\narrowtext
where $C_{k,l}$ $(C^{\dag}_{k,l})$ are spinless conduction electrons
with orbital momentum or channel index $l$, and $d$ ($d^{\dag}$) operators
correspond to a localized impurity. As required by symmetry, the localized
impurity hybridizes only with one-channel, $l=0$. The chemical potential is set
to zero, and the model has particle-hole symmetry. It should be noted that this
model Hamiltonian is exactly the spinless version of the generalized
Anderson impurity model [8].

In order to reveal the physics contained in this model, we first
separate the model Hamiltonian into two part, $H=H_h+H_s$,
where
%\widetext
 \begin{equation}\label{h_h}
H_h= \sum_{k}\epsilon_k C_{k,0}^{\dag}C_{k,0} + \frac{t}{\sqrt{N}}
  \sum_k (C^{\dag}_{k,0}d + d^{\dag}C_{k,0}) + \frac{V_0}{N}\sum_{k,k'}
   C^{\dag}_{k,0}C_{k',0}( d^{\dag}d-\frac{1}{2}),
\end{equation}
\begin{equation}\label{h_s}
H_s= \sum_{k,l\neq 0}\epsilon_k C_{k,l}^{\dag}C_{k,l} + \frac{1}{N}
   \sum_{k,k',l\neq 0}V_l C^{\dag}_{k,l}C_{k',l}
   (d^{\dag}d-\frac{1}{2}).
\end{equation}
%\narrowtext
The hybridizing part $H_h$ is just the single-channel resoant-level
model, corresponding to the single-channel Kondo problem [12,13], while
the screening part
$H_s$ is the multi-channel X-ray edge problem [5]. When
these two parts are coupled through the impurity, there appears an interesting
competition between these two different kinds of metallic behavior.

First, we deal with $H_s$ by  bosonization [14]. If
only s-wave scattering is concerned, this model can be reduced to a
one-dimensional problem with one Fermi point for each screening channel.
We introduce the density operators
$$\rho_l(k)= \frac{1}{\sqrt{N}}\sum_{q=0}^{k_D-k} C_{q,l}^{\dag}C_{q+k,l},
  \hspace{.5cm} \rho_l(-k)= \frac{1}{\sqrt{N}}\sum_{q=k}^{k_D}
         C_{q,l}^{\dag}C_{q-k,l},
  \hspace{.5cm} k\geq 0, $$
and consider a band of width $k_D$. The electron energies are given by
$ \epsilon_k=(k-k_F)/\rho, $
where $\rho\equiv (hv_F)^{-1}$ is the density of states at the Fermi point.
We can define bosonic operators
$ b_{k,l}=\frac{1}{\sqrt{k}}\rho_l(k)$,
$b_{k,l}^{\dag}=\frac{1}{\sqrt{k}}\rho_l(-k),$
which obey the standard commutation relations.
Thus, the screening part $H_s$ can be transformed into
%\widetext
\begin{equation}
H_s=\sum_{k,l>0}\frac{k}{\rho}b_{k,l}^{\dag}b_{k,l}+ \sum_{k,l>0}
\frac{V_l}{\sqrt{N}}\sqrt{k}(b_{k,l}+b_{k,l}^{\dag})(d^{\dag}d-\frac{1}{2}),
\end{equation}
%\narrowtext
and it can be further diagonalized to
$\sum_{k,l>0}\frac{k}{\rho}b_{k,l}^{\dag}b_{k,l}$ through the canonical
transformation
$$U=exp\{\sum_{k,l>0}\frac{\rho
V_l}{\sqrt{kN}}(b_{k,l}-b_{k,l}^{\dag})(d^{\dag}d-\frac{1}{2})\}.$$
In the mean time, the above transformation also transfers
the singularities of the screening channels into the hybridizing
part via the localized impurity. The model Hamiltonian thus becomes
%\widetext
\begin{equation}\label{H}
 H= \sum_{k}\epsilon_k C_{k}^{\dag}C_{k} + \frac{1}{\sqrt{N}}
  \sum_k (\Delta^{\dag} C^{\dag}_{k}d + \Delta d^{\dag}C_{k}) +
\frac{V_0}{N}\sum_{k,k'}C^{\dag}_{k}C_{k'}(d^{\dag}d-\frac{1}{2})+
\sum_{k,l>0}\frac{k}{\rho}b_{k,l}^{\dag}b_{k,l},
\end{equation}
%\narrowtext
with
$ \Delta \equiv t e^{\{-\sum_{k,l>0}\frac{\rho
V_l}{\sqrt{kN}}(b_{k,l}-b_{k,l}^{\dag})\}}.$
Here we have omitted the index "$0$" for the hybridizing channel. This
is a modified single channel resonant-level model with an effective
hybridization
strength $\Delta$, which can change the behavior of the model
from FL with canonical exponents to non-FL with singular nonuniversal
exponents. This resulting Hamiltonian is exact and includes
all physical features of the MCRL model.

Next, we derive the partition function of the Hamiltonian
(\ref{H}). As in the Kondo problem [17,18], this
Hamiltonian is divided as:
$ H=H_0+H_I, $
%\widetext
\begin{equation}
 H_0= \sum_{k}\epsilon_k C_{k}^{\dag}C_{k} +
 \frac{V_0}{N}\sum_{k,k'}C^{\dag}_{k}C_{k'}(d^{\dag}d-\frac{1}{2})
 +\sum_{k,l>0}\frac{k}{\rho}b_{k,l}^{\dag}b_{k,l},
\end{equation}
%\narrowtext
\begin{equation}
H_I=\frac{1}{\sqrt{N}}
  \sum_k (\Delta^{\dag} C^{\dag}_{k}d + \Delta d^{\dag}C_{k}).
\end{equation}
$H_0$ describes a two-state system where the impurity is
either empty or filled, while $H_I$ plays the role of a dipole operator
causing transitions between these two states. The
spinless fermion opators $d$,$d^{\dag}$ can be exactly mapped onto Pauli
matrices as $\sigma^z=2d^{\dag}d-1$, $\sigma^+=d^{\dag}$, $\sigma^-=d$.
For the partition function $Z$ in the interaction representation (with
$\beta$ as the inverse temperature), we have
$ Z= Tr \left ( e^{-\beta
H_0}Te^{-\int_{0}^{\beta}H_I(\tau)d\tau}\right )$,
where $ H_I(\tau)=e^{H_0\tau}H_I e^{-H_0\tau} $ and $T$ is the time ordering
operator. The partition function can be expanded as: $Z=\sum_n Z_n$.
We introduce the following notations in the presence
(absence) of the impurity:
%\widetext
\begin{equation}
 H_0^{\pm}= \sum_{k}\epsilon_k C_{k}^{\dag}C_{k} \pm
 \frac{V_0}{2N}\sum_{k,k'}C^{\dag}_{k}C_{k'}
  +\sum_{k,l>0}\frac{k}{\rho}b_{k,l}^{\dag}b_{k,l},
\end{equation}
and $ C_0= \frac{1}{\sqrt{N}}\sum_k C_k.$ When performing tracing over
impurity's configurations, we find that $Z_n$ vanishes for odd $n$, while
for even $n$ we obtain $ Z_n =Z_n^{'}+Z_n^{''}$ with
\begin{equation}
 Z_n^{'} = \int_0^{\beta}{d\tau_n}\int_0^{\tau_n-\tau_0} {d\tau_{n-1}}....
   \int_0^{\tau_2-\tau_0}{d\tau_1}
 \langle e^{-(\beta-\tau_n)H_0^{+}}(\Delta
C_0)e^{-(\tau_n-\tau_{n-1})H_0^{-}}(\Delta^{\dag}C_0^{\dag})....
(\Delta^{\dag} C_0^{\dag})e^{-\tau_1 H_0^{+}}\rangle.
\end{equation}
%\narrowtext
The expression for $Z_n^{''}$ is obtained by interchanging
$H_{0}^{+}\leftrightarrow H_{0}^{-}$, $\Delta^{\dag}
C_{0}^{\dag}\leftrightarrow\Delta C_0$. Because of the
rotational invariance, we have  $Z_n^{'}= Z_n^{''}$.

As for the screening channels, the bosonization
can be used for hybridizing electrons as well, so far only s-wave
scattering is involved. The boson operators are
defined as
$$ a_{k}=\frac{1}{\sqrt{kN}}\sum_{q=0}^{k_D-k} C_{q}^{\dag}C_{q+k},
 \hspace{.3cm} a_{k}^{\dag}=\frac{1}{\sqrt{kN}}\sum_{q=k}^{k_D}
         C_{q}^{\dag}C_{q-k},  \hspace{.3cm} k\geq 0. $$
Restricted to the low-lying excitations, the kinetic energy
can also be linearized to yield
%\widetext
\begin{equation}
H_0^{\pm}= \sum_{k>0}\frac{k}{\rho} a_k^{\dag}a_k \pm \frac{V_o}{2\sqrt{N}}
\sum_{k>0}\sqrt{k}(a_k+a_k^{\dag})+
\sum_{k,l>0}\frac{k}{\rho}b_{k,l}^{\dag}b_{k,l},
\end{equation}
%\narrowtext
and $ C_0 = (\sqrt{k_D})
exp\{\sum_{k>0}\frac{1}{\sqrt{kN}}(a_k-a_k^{\dag})\}$.
$H_0^{\pm}$ describes two systems of harmonic oscillators with shifted
origins. There is a similar canonical transformation
$S=exp\{\frac{\rho V_0}{2}\sum_{k>0}\frac{1}{\sqrt{kN}}(a_k-a_k^{\dag})\}$
which connects the shifted oscillators to the unshifted ones.
We then have $ H_0^{+}= S H_f S^{-1}, H_0^{-}=S^{-1}H_fS, $ with
$H_f= \sum_{k>0}\frac{k}{\rho} a_k^{\dag}a_k
+\sum_{k,l>0}\frac{k}{\rho}b_{k,l}^{\dag}b_{k,l} $. Hence
%\widetext
\begin{equation}
  Z_n^{'}=
     \int_0^{\beta}{d\tau_n}\int_0^{\tau_n-\tau_0}{d\tau_{n-1}}......
   \int_0^{\tau_2-\tau_0}{d\tau_1} \langle
   B(i\tau_n)B^{\dag}(i\tau_{n-1})B(i\tau_{n-2})......B^{\dag}(i\tau_{1})
   \rangle,
\end{equation}
%\narrowtext
where $ B(t)= e^{iH_ft}(S^{-1}\Delta C_0 S^{-1})e^{-iH_ft}\equiv
 (t\sqrt{k_D})e^{A(t)} $ and
%\widetext
$$ A(t)=(1-\rho V_0)\sum_{k>0}\frac{1}{\sqrt{kN}}(a_k
e^{i\frac{k}{\rho}t}  -a_k^{\dag}e^{-i\frac{k}{\rho}t})
+ \sum_{k,l>0}\frac{\rho V_l}{\sqrt{kN}}(b_{k,l}e^{i\frac{k}{\rho}t}
-b_{k,l}^{\dag}e^{-i\frac{k}{\rho}t}).$$
Now, we evaluate the thermodynamic expectation values using the
strategies of Schotte [18] to obtain the partition function as
 \begin{equation}\label{coulomb}
Z=2\sum_{n=0}^{\infty}(\tilde{t})^{2n}
\int_0^{\beta} \frac{d\tau_n}{\tau_0}
\int_0^{\tau_n-\tau_0} \frac{d\tau_{n-1}}{\tau_0}......
   \int_0^{\tau_2-\tau_0}\frac{d\tau_1}{\tau_0}
 exp\{(1+\gamma)\sum_{i>j}(-1)^{i+j} ln\mid
\frac{\tau_i-\tau_j}{\tau_0}\mid \},
\end{equation}
%\narrowtext
where the bare dimensionless hybridizing strength $\tilde{t}\equiv
\frac{\rho t}{\sqrt{k_D}}$, the cutoff factor $\tau_0=\frac{\rho}{k_D}$, and
$\gamma\equiv -\frac{2\delta_0}{\pi}+ \sum_l (\frac{\delta_l}{\pi})^2$.
 $\delta_l=\pi\rho V_l$ are the phase shifts in the Born approximation
whose exact values should be $\delta_l=2tan^{-1}(\frac{\pi}{2}\rho V_l)$.
However, for weak interactions $V_l$ they are the same. Moreover, the
total phase shift of the model is
$\delta\equiv\frac{\pi}{2}(1+\sqrt{1+\gamma})$ [12].  The form of the
partition function we obtained is the same as that found by Anderson and Yuval
for the Kondo problem [17]. Comparing our expression and theirs, we
identify $\tilde{t}\leftrightarrow \rho J_{\perp}/2$ and
$(1+\gamma)\leftrightarrow (2-2\rho J_z)$. $(1+\gamma)$ is the interaction
strength between the spin-flips.

We also note that the above partition function for $\gamma=0$ is exactly the
partition function of the following Hamiltonian:
\begin{equation}
H_T= \sum_{k}\epsilon_k C_{k}^{\dag}C_{k} + \frac{t}{\sqrt{N}}
  \sum_k (C^{\dag}_{k}d + d^{\dag}C_{k}).
\end{equation}
This is just the Toulouse limit [19] of the MCRL model, which is
exactly the strong-coupling fixed point of this model, displaying a local FL
behavior. When the model is renormalized to this fixed point, its physics can
be described by a single-channel hybridization effect.

To set up the RG flow equations of the model, we can employ the
poor-man's scaling procedure of Anderson {\it et al.} [15] in the
Coulomb gas representation. The RG equations describe
the flow of the hybridizing strength $\tilde{t}$ as the
bandwidth $\frac{1}{\tau}$ is reduced. They are given by
\begin{eqnarray}
&& \frac{d\gamma}{d (ln\tau)}= -4(1+\gamma)(\tilde{t})^2+O(\tilde{t}^4)
\nonumber \\
&& \frac{d\tilde{t}}{d(ln\tau)}=
\frac{(1-\gamma)}{2}\tilde{t}+O(\tilde{t}^3).
\end{eqnarray}
and $ (\gamma -1)^2 -16 (\tilde{t})^2=const. $
These equations were derived by assuming a small fugacity, $\tilde{t}\ll 1$,
{\it i.e.}, a rare gas of spin-flips. In general, the average spacing
between flips is much greater than $\tau$, {\it i.e.}, $\tau\ll
\mid\tau_i-\tau_j\mid$. As in the Kondo problem, we have ignored the
explicit $\tau_0$ dependence of $\tilde{t}$. Our RG flow equations lead to
two different types of behavior:

(i). For $\gamma<1$, or $\tilde{t}\geq \frac{1}{4}(\gamma-1)$ but $\gamma>1$,
the hybridizing strength $\tilde{t}$ will always be relevant. The system is
controlled by the strong-coupling fixed point, displaying FL-type universal
asymptotic behavior. This fixed point is outside the range of
validity of the poor-man's RG approach, but it gives the correct flow
direction. As seen from the partition function, when $\gamma=0$ the system
will reach the unitary limit with the total phase shift $\delta=\pi$,
which is nothing but exactly the Toulouse limit. Therefore, we can
identify the strong-coupling fixed point with this limit.

We notice that the poor-man's scaling part of this model is similar to the
single-channel Kondo problem, but the behavior in the
strong-coupling regime is somehow different. Due to the emergence of new
relevant variables, the Toulouse limit description for the strong-coupling
limit of the Kondo problem is only qualitatively correct [20].

(ii). For $\gamma>1$ and $\tilde{t}<\frac{1}{4}(\gamma-1)$, the
hybridization strength $\tilde{t}$ will be renormalized to zero,
being an irrelevant variable even with repulsive interaction $V_l$. There
exists a weak-coupling fixed line $\tilde{t}=0$. The system exhibits a
power-law decay of  correlation functions with a non-universal exponent,
{\it i.e.}, the same as for the multi-channel X-ray edge problem.
Due to the large number of the screening channels,
the  orthogonality catastrophe effect is now
sufficiently strong to drive the hybridization to zero. As seen from the
expression for $\gamma$,  at least three
channels of  conduction electrons with maximum phase shifts $\pi$ is
needed  to observe this effect.  In this
regime, the poor-man's scaling  can be justified, so the results obtained
are reliable and self-consistent.

The flow diagram is shown in Fig.1. When the bare
hybridizing strength sits at a particular line,
$\tilde{t}=\frac{1}{4}(\gamma-1)$, the system is in a
marginal state: in the short-time (high-energy) limit, we obtain a
non-FL singular scaling exponents, while in the long-time (low-energy)
limit, a FL behavior will be observed. The system exhibits a nontrivial
crossover between the FL and non-FL exponents. This means
 $\tilde{t}=\frac{1}{4}(\gamma-1)$ corresponds to the FL--non-FL
separation line. Hence, we have given a complete physical picture of the
FL--non-FL transition in the MCRL model, and
have found that this model truly includes
the basic physics of the generalized Anderson impurity
model in the symmetric case [8].

{}From the above asymptotically exact solution of the MCRL model, we can see
that the change of physics at $\gamma=1$ pointed out in Ref.[10] is basically
correct. However, instead of studying the
MCRL model itself, these authors considered an anisotropic multi-channel
Kondo-like model, where the physics of competing X-ray edge singularities
and Kondo behavior is not transparent. There is a crucial difference
between the MCRL model and the multi-channel Kondo model. We would
like to emphasize that the non-FL behavior in the MCRL model is caused by
the X-ray edge singularity appearing in the weak coupling regime, while the
physics of the two (or more) channel Kondo problem is the exhibition of a
non-FL behavior in the strong coupling regime even after discarding the
Anderson orthogonality catastrophe [7]. The origins of these two types of
non-FL behavior might be different.

 In conclusion, we have found an asymptotically exact solution of the
MCRL model, and have given an explicit physical picture
of the FL--non-FL crossover in this model.

{\it Acknowledgments}. G. M. Zhang is indebted to X. Y. Zhang for his many
useful discussion and a earlier collaboration on the related subject.
This work was supported in part by the Chinese Science Foundation for Young
Scientists through Grant No. 19204010.

\newpage
{\bf Figure Caption}
\vskip 1cm
{Fig.1  The RG flow diagram for the multi-channel resonant-level
model derived from the scaling theory of Anderson {\it et al.}

\newpage
% GNUPLOT: LaTeX picture for Fig.1
\setlength{\unitlength}{0.240900pt}
\ifx\plotpoint\undefined\newsavebox{\plotpoint}\fi
\sbox{\plotpoint}{\rule[-0.175pt]{0.350pt}{0.350pt}}%


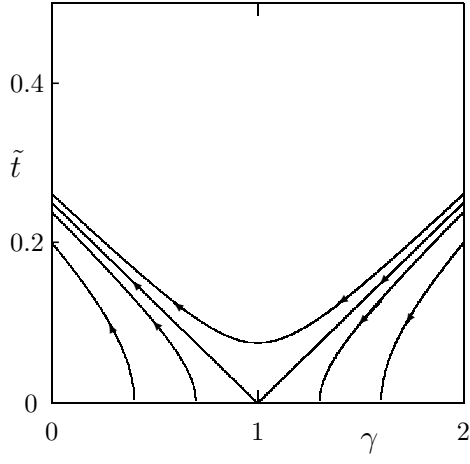
\end{document}